\documentclass[11pt]{llncs}

 \usepackage{amsmath,amssymb}
\usepackage[mathscr]{eucal}
\def\>{\rangle} \def\<{\langle} \def\d{{\rm d}} \def\Cmplx{{\mathbb
    C}} \def\spc#1{{\mathcal #1}} 
\def\map#1{{\mathcal #1}} \def\Tr{{\rm Tr}} 
\begin{document}
%
\title{On quantum estimation, quantum cloning and finite quantum de Finetti theorems}
\author{Giulio Chiribella \inst{1} }
\institute{Perimeter Institute for Theoretical Physics, 31 Caroline Street North, Waterloo, Ontario  N2L 2Y5, Canada.}

\maketitle       
\begin{abstract}
This paper presents a series of results on the interplay between quantum estimation, cloning and finite de Finetti theorems.  
First, we consider the measure-and-prepare channel that uses optimal estimation to convert $M$ copies into $k$ approximate copies of an unknown pure state and we show that  this channel is equal to a random loss of all but $s$ particles followed by cloning from $s$ to $k$ copies. When the number $k$ of output copies  is large with respect to the number $M$ of input copies the measure-and-prepare channel converges in diamond norm to the optimal universal cloning.   In the opposite case, when  $M$ is large compared to $k$, the estimation becomes almost perfect and the measure-and-prepare channel converges in diamond norm to the partial trace over all but $k$ systems.  This result is then used to derive de Finetti-type results for quantum states and for symmetric broadcast channels, that is, channels that distribute quantum information to many receivers in a permutationally invariant fashion.   Applications of the finite de Finetti theorem for symmetric broadcast channels include  the derivation of diamond-norm bounds on the asymptotic convergence of quantum cloning to state estimation and the derivation of bounds on the amount of quantum information that can be jointly decoded by a group of $k$ receivers at the output of a symmetric broadcast channel.
\end{abstract}

The connection between quantum estimation and cloning is an inspiring leitmotiv of Quantum Information Theory \cite{gismass,wernerclon,bem98,kwern,cinche,fasegiusta,bae,me}.  The main related question is:  how well can we simulate  cloning via estimation?  Or, more precisely, how well can we simulate cloning  with a ``measure-and-prepare"  protocol where the input systems are measured, and the output systems are prepared in some state depending on the measurement outcome?  As a particular instance of  this question, one can ask whether  ``asymptotic cloning is state estimation"  \cite{keylprob}, that is, whether the gap between the single-particle fidelity of an optimal cloning channel and the fidelity of the corresponding optimal estimation  vanishes when the number of clones tends to infinity.   

In Ref. \cite{bae}  Bae and Ac\'in showed that a channel producing an infinite number of indistinguishable clones must be of the measure-and-prepare form.   On the other hand, Ref. \cite{me}  showed that a  channel producing a finite number $M<\infty$ of  indistinguishable clones can be simulated by a measure-and-prepare channel introducing an error at most of order $\mathcal O(1/M)$ on each clone.   The proof of Ref. \cite{me} was based on the so-called \emph{finite quantum de Finetti theorem} \cite{oneandhalf,renner,facile}, that states that the restriction to $k$ particles  of a permutationally invariant $M$-partite state can be approximated with an error at most of order $\mathcal O(k/M)$ by a mixture of product states of the form $\rho^{\otimes k}$.   This theorem represents the finite version of the \emph{quantum de Finetti theorem} proved by Caves, Fuchs, and Schack \cite{qdefi} in the context of the Bayesian interpretation of quantum theory.  The quantum de Finetti theorem of Ref. \cite{qdefi}  corresponds to the ideal  $M = \infty$ case and can be directly seen as the quantum formulation of the celebrated de Finetti theorem \cite{defi}.

Apparently, finite quantum de Finetti theorems are the key to prove the equivalence between asymptotic cloning and estimation. The first result of this paper is to show that, in a sense, the converse is also true:  a finite quantum de Finetti-type result can be derived from a particular relation between the optimal estimation  \cite{maspop,bem98} and the optimal cloning \cite{wernerclon} of an unknown pure state.   Precisely, we will see that the optimal measure-and-prepare channel sending $M$ copies of an unknown pure states to $k$ approximate copies is equivalent to a random loss of all but $s$ particles followed by universal cloning from $s$ to $k$ copies. For $M >> k$  the term with $s=k$ dominates,  implying that the optimal measure-and-prepare channel is close to the partial trace over  all but $k$ particles. As we will see, this implies directly a de Finetti-type result.   Qualitatively, this result shows that the working principle of the finite de Finetti theorems is simply the fact that state estimation from $M$ input copies to $k$ output copies becomes almost perfect when $M$ is large compared to $k$. Quantitatively, however, the bound derived from the representation of the optimal measure-and-prepare channel as a random mixture of losses followed by cloning can be tightened, as mentioned in subsection \ref{sec:improve}.  The bound can be used to derive a finite de Finetti theorem for symmetric quantum broadcast channels, i.e. for channels that distribute quantum information to $M$ indistinguishable users.  Examples of symmetric broadcast channels are the channels for the optimal cloning of an unknown state $\rho_i$ randomly drawn with probability $p_i$ from some set of states $\{\rho_i\}$ \cite{clonrev}. 
The paper concludes with two applications of the finite de Finetti theorem for symmetric broadcast channels. First, the theorem will be used to
 provide  diamond-norm bounds on the asymptotic convergence of quantum cloning to state estimation, thus strengthening the proof of Ref. \cite{me}. As a second application, the theorem will be used to show that  the restriction to $k$ users of any symmetric broadcast channel  has a quantum capacity that vanishes at rate $\mathcal O (k/M)$  in the large $M$ asymptotics. Even if the overall channel is unitary, and therefore its capacity has the maximum possible value,  a group of $k<<M$ users will only be able to decode a vanishingly small amount of quantum information.

\section{The universal measure-and-prepare channel}\label{sec:univmp}
Let us start with some simple facts about the optimal measure-and-prepare channel transforming $M$ copies of a completely unknown pure states into $k$ approximate copies.  
The optimal quantum measurement for the estimation of a completely
unknown pure state $|\psi\> \in \spc H \simeq \Cmplx^d$ from $M$ input copies is given by the \emph{coherent-state POVM}  \cite{maspop,bem98}
\begin{equation}\label{cohpovm}
  P^{(M)}_\varphi ~\d \varphi  =  d^{(M)}_+~ |\varphi \>\< \varphi|^{\otimes M} ~ \d \varphi \qquad d^{(M)}_+ = \begin{pmatrix}  d+ M-1 \\
    M  \end{pmatrix}
\end{equation}
where $|\varphi\>\in\spc H$ is a unit vector and $\d \varphi$ is the
normalised $SU(d)$-invariant measure on pure states.  This measurement
provides a resolution of the identity in the symmetric subspace
$\left(\spc H^{\otimes M}\right)_+ \subseteq \spc H^{\otimes M}$,
namely in the subspace spanned by the unit vectors
\begin{equation}\label{symbasis}
|\vec n\> :=   \frac 1 {\sqrt{M!  n_1! n_2! \dots n_d!}} \sum_{ \pi \in S_M} U^{(M)}_\pi |1\>^{\otimes n_1} |2\>^{\otimes n_2} \dots |d\>^{\otimes n_{d}}
\end{equation}
where $|1\>,|2\>, \dots, |d\>$ is a fixed orthonormal basis for $\spc
H$, $\vec n = (n_1, n_2,\dots, n_d) $ is a partition of $M$, the sum
runs over the symmetric group $S_M$ of all permutations of $M$ objects, and $U^{(M)}_\pi$ is the
unitary operator that permutes the $M$ copies of $\spc H$ according to the permutation $\pi \in S_M$.

Denoting by $\spc P_{M,d}$ the set of partitions of $M$ in $d$ nonnegative integers, the normalization of the coherent-state POVM in Eq. (\ref{cohpovm}) is given by
\begin{equation}\label{normpovm}
  \int \d \varphi ~ P^{(M)}_\varphi = \sum_{\vec n \in \spc P_{M,d}}  |\vec n\>\<\vec n| = P^{(M)}_+,   
\end{equation}
where $P^{(M)}_+$ is the projector on the symmetric subspace $(\spc
H^{\otimes M})_+$. 

We now consider the \emph{universal measure-and-prepare channel} from
$M$ to $k$ copies, namely the channel that measures the coherent-state
POVM $P_\varphi^{(M)}$ and, according to the estimate, prepares $k$ copies of the state
$|\varphi\>$:
\begin{equation}
  \map {UMeasPrep}_{M,k} (\rho) := \int \d \varphi  ~\Tr[ P^{(M)}_\varphi  \rho] ~ |\varphi\>\<\varphi|^{\otimes k}.
\end{equation}
Using Eq. (\ref{normpovm}) with the substitution $M \to M+k$ one
obtains the equivalent expression
\begin{equation}\label{canaleproiettore}
\begin{split}
  \map {UMeasPrep}_{M,k} (\rho) & = d^{(M)}_+ ~\int \d \varphi ~ \Tr_M \left[\left(\rho \otimes I^{\otimes k} \right)  |\varphi\>\<\varphi|^{\otimes M+k}\right]\\
  & = \frac{ d_+^{(M)}}{d_+^{(M+k)}} \Tr_{M}\left[\left(\rho \otimes
      I^{\otimes k} \right) P^{(M+k)}_+\right]
\end{split}  
\end{equation}
where $\Tr_M$ denotes the partial trace over the first $M$ Hilbert
spaces.

For an arbitrary pure state $|\psi\>$, the fidelity between the
channel output $\map {UMeasPrep}_{M,k}  (|\psi\>\<\psi|^{\otimes M})$ and the desideratum $|\psi\>\<\psi|^{\otimes k}$ is given by
$F_{M,k} = d_+^{(M)}/d_+^{(M+k)}$,  as it
is immediate from Eq. (\ref{canaleproiettore}). In fact,  it is easy to show that $F_{M,k} =
d_+^{(M)}/d_+^{(M+k)}$ is the maximum average fidelity achievable with a
measure-and-prepare channel $\map M (\rho) = \sum_i  \Tr[P_i \rho]  \rho_i$, where $\{P_i\}$ is a POVM on $\left( \spc H^{\otimes M} \right)_+$ and $\{\rho_i\}$ is a set of states on $\left( \spc H^{\otimes k}\right)_+$. Indeed, in this case one has 
\begin{align}
\overline F  &= \int \d \psi  \<  \psi|^{\otimes k}  \map M\left(|\psi\>\<\psi|^{\otimes M} \right) |\psi\>^{\otimes k} = \frac{\sum_i  \Tr \left[(P_i \otimes \rho_i)  P^{(M+k)}_+\right]}{d_+^{(M+k)}}  \nonumber \\
&\le   \frac{\sum_i  \Tr \left[ P_i \otimes \rho_i \right]}{d_+^{(M+k)}}  = \frac{d_+^{(M)}} {d^{(M+k)}_+} \nonumber
\end{align} 
(cf. Bru\ss ~and Macchiavello
\cite{bem98} for the $k=1$ case).  Clearly, when $M$ is large compared to $k$
the fidelity $F_{M,k}$ is close to unit: the desired output states
$|\psi\>^{\otimes k}$ are much less distinguishable than the input
states $|\psi\>^{\otimes M}$, thus allowing for an almost ideal
re-preparation. In this case, one has 
\begin{equation*}
\map {UMeasPrep}_{M,k}  \left(|\psi\>\<\psi|^{\otimes M}\right)  \approx |\psi\>\<\psi|^{\otimes k} \qquad \forall |\psi\>  \in \spc H ,
\end{equation*}
or, equivalently (cf. the Appendix), 
\begin{equation*}
\map {UMeasPrep}_{M,k}  (\rho)  \approx \Tr_{M-k} [\rho] \qquad \forall \rho \in {\sf Lin} \left(   \left( \spc H^{\otimes M}\right)_+ \right), 
\end{equation*}
where ${\sf Lin} (V)$ denotes the set of linear operators on the linear space $V$ ($V = \left( \spc H^{\otimes M}\right)_+$ in this case).  
 Despite the simplicity of the above observation, the
consequences of the fact that for $M >>k$ the estimation from $M$ to $k$ copies is
``almost ideal'' are far from trivial: as we will see, this simple fact can be considered as the working principle of the finite de Finetti theorems. 

The purpose of the next subsection is to give a convenient
representation of the channel $\map {UMeasPrep}_{M,k}$ as a convex
mixture of losses concatenated with cloning channels.  Using this representation we
will show that in the limit $k/M\to 0$ the channel $\map
{UMeasPrep}_{M,k}$ converges to the partial trace $\Tr_{M-k}$ in the strongest
possible sense, in terms of the  \emph{diamond norm} \cite{diamond}, equivalent to the \emph{norm of complete boundedness}
\cite{paulsen} of the channel in Heisenberg picture.  Operationally, convergence in the diamond norm means that for $M>>k$ the two channels $\map
{UMeasPrep}_{M,k}$ and $\Tr_{M-k}$ are almost indistinguishable even
when entanglement-assisted discrimination strategies are employed. 

\subsection{Representation of the universal measure-and-prepare channel as a mixture of universal cloning channels}\label{sec:representation}

The main result of this subsection is the following expression, proved in the Appendix: 
\begin{equation}\label{main}
  \map {UMeasPrep}_{M,k} (\rho)  =  \sum_{s=0}^{\min\{k,M\}}  p_s ~ \map {UClon}_{s,k} \left(\Tr_{M-s}[\rho]\right), \quad p_s = \frac{\begin{pmatrix}  M \\ s \end{pmatrix} \begin{pmatrix}  d+k-1 \\ k-s \end{pmatrix}}{\begin{pmatrix}  d+M +k-1 \\ k \end{pmatrix}}, 
\end{equation}
$\map{UClon}_{s,k}$ being the \emph{universal $s$-to-$k$ cloning channel}, i.e. the optimal quantum channel that clones an unknokwn pure state
$|\psi\>$ from $s$ to $k$ copies, given by \cite{wernerclon,kwern}
\begin{equation}\label{clon}
  \map {UClon}_{s,k} (\rho) = \frac{d^{(s)}_+}{d^{(k)}_+} ~ P^{(k)}_+  \left (\rho \otimes I^{\otimes (k-s)} \right) P^{(k)}_+.   
\end{equation}
Note that $\{p_s\}$ is a probability distribution, as  the normalization  
\begin{equation*}
\sum_{s=0}^{\min\{k,M\}} p_s =\sum_{s=0}^k p_s =1
\end{equation*} 
follows immediately from the fact that $p_s = 0$ if $s>M$ and from the Chu-Vandermonde convolution formula (see Eq. (7.6) p. 59  of Ref. \cite{askey} for an equivalent formula)
\begin{equation}\label{chuvander}
\begin{pmatrix}    z+ w  \\  N\end{pmatrix}  =  \sum_{i=0}^N  \begin{pmatrix}    z  \\  i\end{pmatrix}  \begin{pmatrix}    w  \\   N-i \end{pmatrix}  \qquad \forall z,w \in \Cmplx, \forall N \in \mathbb{N} .
\end{equation}
Eq. (\ref{main}) means that measuring $M$ copies and re-preparing $k$ copies has the same effect of a random loss of $M-s$ systems followed by quantum cloning from $s$ to $k$ copies: the particles that are missing are replaced by clones. 

In the following we will consider the two extreme cases $k>> M$ and $M>>k$. In the former, we will see that the measure-and-prepare channel $\map {UMeasPrep}_{M,k}$ converges to the universal cloning $\map {UClon}_{M,k}$.  In the latter, the measure-and-prepare channel $\map {UMeasPrep}_{M,k}$ will converge to the partial trace $\Tr_{M-k}$, leading to a de Finetti-type result.  The convergence will be quantified in terms of the \emph{diamond norm} \cite{diamond}  (in Heisenberg picture, the completely bounded norm \cite{paulsen}), which for a Hermitian-preserving map $\Delta$ from ${\sf Lin}(\spc H_{in})$ to ${\sf Lin} ( \spc H_{out})$ is given by
\begin{equation}
|\!|  \Delta  |\!|_\diamond =  \sup_{\spc H_A} \sup_{|\Psi\>  \in \spc H_A \otimes \spc H_{in}, |\!|  \Psi|\!| =1}    |\!|(\map I_A \otimes \Delta) (|\Psi\>\<\Psi|)|\!|_1,
\end{equation}
where $|\!|  A |\!|_1 = \Tr |A|$ is the trace-norm and $\map I_A$ is the identity map on the ancillary Hilbert space $\spc H_A$.

\subsection{$k>>M$ case: convergence to universal cloning}
Suppose that the number of output copies $k$ is larger than the number of input copies $M$. In the limit of  $M/k \to 0$, the term with $s=M$ in Eq. (\ref{main}) dominates, thus giving $\map{UMeasPrep}_{M,k}  \approx  \map {UClon}_{M,k}$.  
 
An estimate of the diamond-norm convergence  to universal cloning is given by the following:
\begin{theorem}[Convergence to universal cloning]\label{superconvergence}
The universal measure-and-prepare channel $\map{UMeasPrep}_{M,k}$ converges to the universal cloning channel $\map{UClon}_{M,k}$ in the limit $k \to \infty$. In particular, the following bound holds:  
\begin{equation}
\left|\!\left|   \map {UMeasPrep}_{M,k}  -  \map {UClon}_{M,k} \right|\!\right|_{\diamond} \le \frac{2M(d+M-1)}{k+d}.   
\end{equation} 
\end{theorem}
{\bf Proof.} Writing $\map {UMeasPrep_{M,k}} = p_M \map{UClon}_{M,k}  + (1-p_M) \map {Rest}$ where $\map {Rest}$ is a suitable channel, one has $|\!|  \map {UMeasPrep}_{M,k}  -  \map{UClon}_{M,k}     |\!|_\diamond \le (1-p_M)   |\!|  \map{Rest} - \map {UClon}_{M,k} |\!|_\diamond$.   Since the distance between the two channels $\map {Rest}$ and $\map{UClon}_{M,k}$ is upper bounded by $2$, this gives $|\!|  \map {UMeasPrep}_{M,k} - \map{UClon}_{M,k}|\!|_\diamond \le 2 (1-p_M)$.  The bound in Eq. (\ref{bound1}) just comes from a lower bound on $p_M$:  
\begin{equation}
\begin{split}\nonumber
p_M &= \frac{k (k-1) \dots (k-M+1)}{(d+M+k-1) (d+M+k-2) \dots (d+k)} \ge \left ( \frac{ k-M+1}{d+k}\right)^M \\
& = \left ( 1- \frac{d+M-1}{d+k}\right)^M \ge 1-   \frac{  M(d+M-1)}{d+k}.
\end{split}
\end{equation}
\qed    

Theorem \ref{superconvergence} shows an exceptionally strong case of equivalence between asymptotic cloning and state estimation: it shows that, in the universal case, the optimal cloning channel  \cite{wernerclon,kwern} converges in diamond norm to the measure-and-prepare channel $\map {UMeasPrep}_{M,k}$ when the  number $k$ of output copies is large with respect to the number $M$ of input copies.   It is worth stressing, however, that this result is very specific to the universal case.  What can be proved for generic (i.e. non-universal) cloning channels is that the  $k$-particle restrictions of a cloning channel with $M$ output copies can be simulated by  a measure-and-prepare channel with an error of order $k/M$ (see subsection \ref{subsec:strongeq}).  This result will emerge from the analysis  of Eq. (\ref{main}) in  the $M>>k$ case,  which is discussed in the next subsection.

\subsection{$M>>k$ case: convergence to the partial trace}
Here we consider the case where the number is input copies $k$ is large with respect to the number of output copies $M$.  In this case, the leading term in Eq. (\ref{main}) is the term with $s=k$. Note that, since for $s=k$ the universal cloning $\map {UClon}_{k,k}$ is simply the identity map
on $(\spc H^{\otimes k})_+$,  the corresponding term in
Eq. (\ref{main}) is the partial trace $\Tr_{M-k}$. 
Therefore,  when $M$ is large compared to $k$ the channel $\map{UMeasPrep}_{M,k}$ converges to the trace
$\Tr_{M-k}$. This implies an almost ideal estimation, with
$\map{UMeasPrep}_{M,k} (|\psi\>\<\psi|^{\otimes M}) \approx \Tr_{M-k} [  |\psi\>\<\psi|^{\otimes M}] =
|\psi\>\<\psi|^{\otimes k}$.   A first estimate on the diamond-norm convergence to ideal estimation is given by the following
\begin{theorem}[Convergence to ideal estimation]
The universal measure-and-prepare channel $\map{UMeasPrep}_{M,k}$ converges to the trace channel $\Tr_{M-k}$ in the limit $M\to \infty$. In particular, the following bound holds
\begin{equation}\label{bound1}
\left|\!\left|  \map{UMeasPrep}_{M,k}  - \Tr_{M-k}  \right|\!\right|_\diamond  \le  \frac{ 2 k(d+k-1)}{M+d}.
\end{equation}   
\end{theorem}
{\bf Proof.} Writing $\map {UMeasPrep_{M,k}} = p_k \Tr_{M-k}  + (1-p_k) \map {Rest}$ where $\map {Rest}$ is a suitable channel, one has $|\!|  \map {UMeasPrep}_{M,k}  - \Tr_{M-k}    |\!|_\diamond \le (1-p_k)   |\!|  \map{Rest} - \Tr_{M-k} |\!|_\diamond$.   Since the distance between the two channels $\map {Rest}$ and $\Tr_{M-k}$ is upper bounded by $2$, this gives $|\!|  \map {UMeasPrep}_{M,k} - \Tr_{M-k}|\!|_\diamond \le 2 (1-p_k)$.  The bound in Eq. (\ref{bound1}) just comes from a lower bound on $p_k$:  
\begin{equation}
\begin{split}\nonumber
p_k &= \frac{M (M-1) \dots (M-k+1)}{(d+M+k-1) (d+M+k-2) \dots (d+M)} \ge \left ( \frac{ M-k+1}{d+M}\right)^k \\
& = \left ( 1- \frac{d+k-1}{d+M}\right)^k \ge 1-   \frac{  k(d+k-1)}{d+M}.
\end{split}
\end{equation}
\qed
The bound of Eq. (\ref{bound1}) clearly implies a de Finetti-type result:  
\begin{corollary}
For every state  $\rho$ with support in the symmetric space $(\spc H^{\otimes M})_+$ there exists a state
$\tilde \rho = \sum_i p_i  |\psi_i\>\<\psi_i |^{\otimes M}$ such that  the $k$-particle restrictions of $\rho$ and $\tilde \rho$ are almost indistinguishable for large $M$. Precisely, denoting the $k$-particle restrictions by $\rho^{(k)} = \Tr_{M-k} [\rho]$ and $\tilde \rho^{(k)} =
\Tr_{M-k} [\tilde \rho]$, one has 
\begin{equation}\label{define}
\left|\!\left|  \rho^{(k)}  -\tilde \rho^{(k)} \right|\!\right|_1 \le  \frac{ 2 k(d+k-1)}{M+d}
\end{equation}   
\end{corollary}
{\bf Proof.} Taking $\tilde \rho =
\map{UMeasPrep}_{M,M} (\rho)$ we obtain a state of the desired form, and, in addition, we have
\begin{equation}\label{definetti}
\begin{split}\nonumber
  \left \| \tilde \rho^{(k)} -\rho^{(k)} \right\|_1 & = \left \| \map{UMeasPrep}_{M,k}
  (\rho)-\Tr_{M-k} [\rho] \right\|_1 \\
  & \le \left|\!\left|  \map{UMeasPrep}_{M,k}  - \Tr_{M-k}  \right|\!\right|_\diamond  \\
  &\le \frac{ 2 k(d+k-1)}{M+d}.
\end{split}
\end{equation}  \qed
The bound of Eq. (\ref{define}) can be extended  to the case of 
states on $\spc H^{\otimes M}$ that are just permutationally
invariant, using the fact that \emph{i)} every permutationally invariant state
on $\spc H^{\otimes M}$ has a purification in the
symmetric space $(\spc K^{\otimes M})_+$, with $\spc K = \spc H
\otimes \spc H$ (see e.g. \cite{oneandhalf}) and that \emph{ii)}  the norm is non-increasing under partial traces.  Therefore, for a
permutationally invariant state the bound of Eq. (\ref{define}) holds with the substitution $d \to
d^2$.


\subsection{Improving the bound}\label{sec:improve}  
The bound of Eq. (\ref{bound1}) provides good estimates for $k=1$ or when $d$ is large, so that $Mk \le d^2$ (see the observation below).    Outside this range of values, 
the estimate can be improved using the technique developed in Ref. \cite{oneandhalf} for the proof of the finite de Finetti theorem, combined with the bounding of Ref. \cite{me}: 
\begin{theorem}
The universal measure-and-prepare channel $\map {UMeasPrep}_{M,k}$ satisfies the bound
\begin{equation}\label{bound2}
|\!|  \map {UMeasPrep}_{M,k} - \Tr_{M-k}  |\!|_\diamond \le 4 \left ( 1 - \sqrt{\frac{d_+^{(M-k)}} {d_+^{(M)}} }  \right) \le    \frac{2kd}M 
\end{equation}
\end{theorem}
\emph{Observation.} Note that the quantity $2kd/M$ in Eq. (\ref{bound2}) is larger than the quantity $2k(d+k-1)/(M+d)$ in Eq. (\ref{bound1}) whenever $M(k-1) \le d^2$. 
In general, the more accurate estimate is obtained by taking the minimum between the two quantities in Eqs. (\ref{bound1}) and (\ref{bound2}). 
\medskip

{\bf Proof of Theorem 2.}  Let $|\Psi\>$ be an arbitrary state in $\spc H_A \otimes \left(\spc H^{\otimes M}\right)_+$, where $\spc H_A$ is an arbitrary Hilbert space.  Define the states
\begin{equation*} 
\begin{split}
\rho^{(Ak)}  &= \left(\map I_A\otimes \Tr_{M-k} \right) [|\Psi\>\<\Psi|]\\
 \tilde \rho^{(Ak)}  &=\left(  \map I_A\otimes \map {UMeasPrep}_{M,k} \right) [|\Psi\>\<\Psi|].
 \end{split}
 \end{equation*} 
Using the normalization of the coherent-state POVM in Eq. \eqref{normpovm} with the substitution $M \to M-k$, we  can write $\rho^{(Ak)} =  
\int \d \varphi~ \rho^{(Ak)}_\varphi$, where
\begin{equation*} 
\rho^{(Ak)}_\varphi=  \Tr_{M-k}  \left[|\Psi\>\<\Psi| \left( I_A
\otimes I^{\otimes k} \otimes P^{(M-k)}_\varphi\right) \right]~.
\end{equation*}   
On the other hand, the state $\widetilde \rho^{(Ak)}$ can be
written as
\begin{equation*} 
\widetilde \rho^{(Ak)} = \lambda \int \d
\varphi~ \left( I_A \otimes P^{(k)}_\varphi \right) ~ \rho^{(Ak)}_\varphi ~
\left( I_A \otimes P^{(k)}_\varphi \right), 
\end{equation*} 
with $\lambda =  \frac {d^{(M)}_+}{d^{(M-k)}_+ d^{(k) 2}_+}$.  
The difference between $\rho^{(Ak)}-\tilde \rho^{(Ak)}$ is then
given by
\begin{equation*}
\rho^{(Ak)} - \widetilde \rho^{(Ak)}=  \int \d \varphi~   \left(A_\varphi - B_\varphi A_\varphi B_\varphi \right),
\end{equation*}
where $A_\varphi=\rho^{(Ak)}_\varphi$ and $B_\varphi= \sqrt{\lambda} \left( I_A \otimes P^{(k)}_\varphi  \right)$.  

Using the relation $A -BAB= A (I -B) + (I -B) A -(I -B) A(I-B)$ 
we obtain 
\begin{equation}\label{Rho-RhoTilde}
\rho^{(Ak)} -\widetilde \rho^{(Ak)} =  C
+ C^\dag -D~,
\end{equation} 
where $C=  \int \d \varphi~ A_\varphi \left( I -B_\varphi \right)$ and  $D = \int \d \varphi~  \left(I -B_\varphi \right) A_\varphi \left( I -B_\varphi \right)$. The operator $C$ can be calculated using the relation       
\begin{equation*}\label{Int}
\begin{split}
\int \d \varphi~ A_\varphi B_\varphi&= \frac{\sqrt{\lambda} d^{(k)}_+ d^{(M-k)}_+}{d^{(M)}_+} ~\int \d \varphi~\Tr_{M-k} \left[|\Psi\>\<\Psi| ~ \left (I_A \otimes P^{(M)}_\varphi \right)\right]\\   
&=  \sqrt{\frac{d^{(M-k)}_+}{d^{(M)}_+}}~ \Tr_{M-k} [|\Psi\>\<\Psi|]=  \sqrt{\frac{ d^{(M-k)}_+}{d^{(M)}_+}}~  \rho^{(Ak)},
\end{split}
\end{equation*}
which gives $C= \left(1-\sqrt{d^{(M-k)}_+ /d^{(M)}_+}\right) \rho^{(Ak)} = C^\dag$.

Taking the norm on both sides of Eq.  \eqref{Rho-RhoTilde}, using the triangle inequality, and the fact that $C$ and $D$ are both nonnegative we obtain $|\!|\rho^{(Ak)} -\widetilde \rho^{(Ak)}|\!|_1 \le  2|\!|C|\!|_1 + |\!|D|\!|_1 = 2 \Tr[C] + \Tr[D]$.   Finally, taking the trace on both sides of Eq. \eqref{Rho-RhoTilde}  we get $\Tr[ D] = 2 \Tr[C] $.  The inequality $|\!| \rho^{(Ak)} -\tilde \rho^{(Ak)}|\!|_1 \le  4 \Tr [C]$  then gives the first bound in Eq. (\ref{bound2}). The second bound follows from the inequalities $d^{(M-k)}_+/d^{(M)}_+ \ge  (1- k/M)^{d}$ (see e.g. Ref.\cite{oneandhalf}) and   $(1-x)^\alpha \ge  1-\alpha x$, which holds for $\alpha\ge 1$ and $x\le 1$. \qed

\section{Symmetric broadcast channels}
A \emph{quantum broadcast channel} is a channel with a single sender and many receivers \cite{broadchan}.   We define a \emph{symmetric} broadcast channel as a channel where the Hilbert spaces of all receivers are isomorphic and the output of the channel is invariant under permutations.  Precisely, we say that a channel $\map E:  \mathsf{Lin}  (\spc H_{in}) \to \mathsf{Lin}  \left(\spc H^{\otimes M}\right)$ is a \emph{symmetric broadcast channel} if
\begin{equation}\label{defsymbroadchan}
\map E =  \map U^{(M)}_{\pi}  \map E \qquad \forall \pi \in S_M,
\end{equation}
where $\map U^{(M)}_\pi$ is the unitary channel defined by $\map U^{(M)}_\pi (\rho)  := U^{(M)}_\pi \rho U^{(M) \dag}_\pi$, $\rho \in \mathsf{Lin}  (\spc H_{in})$.  The requirement of Eq. (\ref{defsymbroadchan}) models the situation where the quantum information in the input is equally spread over all receivers: any possible permutation of the receivers leaves the channel invariant.  
 An example of symmetric broadcast channel is the optimal cloning channel for an arbitrary set of pure states, whenever the figure of merit is the average of the single-copy fidelity over all the $M$ output copies (see e.g. \cite{kwern}).  In the following we will prove a finite de Finetti theorem for symmetric broadcast channels. The theorem is then used to show a strong form of the equivalence between asymptotic cloning and state estimation and to provide  bounds on the amount of quantum information that can be jointly decoded by $k$ receivers at the output of a symmetric broadcast channel.  
\subsection{Finite de Finetti theorems for symmetric quantum broadcast channels}
For symmetric broadcast channels with output in the symmetric subspace the following approximation result holds: 
\begin{theorem}[Finite de Finetti theorem for symmetric broadcast channels with output in the symmetric subspace]\label{theo:ssymbroad}
For a symmetric broadcast channel $\map E:  {\sf Lin} (\spc H_{in}) \to  {\sf Lin} \left( (\spc H^{\otimes M})_+\right)$ there is a measure-and-prepare channel $\widetilde {\map E}$ of the form $\widetilde{\map E} (\rho)  =  \sum_i \Tr[P_i \rho]  ~ |\psi_i\>\<\psi_i|^{\otimes M}$
such that 
\begin{equation}\label{bound3}
|\!|  \widetilde{\map E}^{(k)}  -\map E^{(k)}   |\!|_\diamond \le 4 \left ( 1 - \sqrt{\frac{d_+^{(M-k)}} {d_+^{(M)}} }  \right) \le    \frac{2kd}M ,
\end{equation}
where $\widetilde{\map E}^{(k)} := \Tr_{M-k} \circ \widetilde{\map E}$ and $\map E^{(k)} := \Tr_{M-k} \circ \map E$.
\end{theorem}

{\bf Proof} Define the measure-and-prepare channel $\widetilde
{\map E}$ as
\begin{equation*}
\widetilde {\map E} (\rho) = \map {UMeasPrep_{M,M}}\circ \map E (\rho) = \int \d \varphi ~  \Tr[Q_{\varphi} \rho] ~ |\varphi\>\<\varphi|^{\otimes M},
\end{equation*}
where $Q_\varphi \d \varphi$ is the POVM defined by
\begin{equation*} \Tr[Q_\varphi \rho] = \Tr[P_\varphi^{(M)} \map E (\rho)]~ \forall \rho\in {\sf Lin} (\spc H_{in}),
\end{equation*}
 that is, $Q_\varphi \d \varphi$ is the POVM obtained by applying the channel $\map E$ in Heisenberg picture to the coherent-state POVM $P^{(M)}_{\varphi} \d \varphi$.  From the definition of $\widetilde {\map E}$ it is clear that $\map E^{(k)} =\map{UMeasPrep}_{M,k}  \circ \map E$.  
Using  the submultiplicativity property  $|\!|\map A \map B|\!|_\diamond \le |\!| \map A|\!|_\diamond |\!| \map B|\!|_\diamond$, the fact that $|\!|  \map E|\!|_\diamond =1 $ since $\map E$ is a channel, and the bound of Eq. (\ref{bound2})  we then obtain
\begin{equation}
\begin{split}\nonumber
 \left \| \widetilde{\map E}^{(k)}-\map E^{(k)} \right\|_\diamond &= \|
  (\map{UMeasPrep_{M,k}} - \Tr_{M-k} )\circ \map E\|_\diamond\\
  & \le 4 \left(1-\sqrt{\frac{d_+^{(M)}}{d_+^{(M+k)}}}\right)\le \frac{2dk}M.
\end{split}
\end{equation} 
 \qed 
 The extension to arbitrary broadcast channels with permutationally invariant output is given in the following
\begin{theorem}[Finite de Finetti theorem for symmetric broadcast channels]\label{theo:symbroad}
For every symmetric broadcast channel $\map E:  {\sf Lin } (\spc H_{in})  \to {\sf Lin} \left( \spc H^{\otimes M}\right)$ there is a measure-and-prepare channel  $\widetilde {\map E} = \sum_{i} \Tr[P_i\rho]  \rho_i^{\otimes M}$ such that  the bounds in Eq. (\ref{bound3}) hold with the substitution $d\to d^2$.
\end{theorem}
{\bf Proof}  Consider the Stinespring dilation $\map E(\rho) =  \Tr_{env} [ V \rho V^\dag]$, where $V: \spc H_{in} \to \spc H^{\otimes M}  \otimes \spc H_{env}$ is an isometry and $\Tr_{env}$ is the partial trace over the environment Hilbert space $\spc H_{env}$.    Since by definition a symmetric broadcast channel satisfies the relation 
\begin{equation*}
\map E (\rho) =  U_\pi^{(M)} \map E(\rho) U_\pi^{(M)}, \qquad \forall \rho\in{\rm Lin} (\spc H_{in}), \forall \pi \in S_M,
\end{equation*}  
it follows from the theory of covariant channels that one can choose $\spc H_{env} = \spc H^{\otimes M} \otimes \spc H_{in} $ and  $V$ with the property
\begin{equation*}
\left(U_\pi^{(M)} \otimes U^{(M)}_\pi  \otimes I_{in}\right) V = V, \qquad \forall \pi \in S_M
\end{equation*} 
(see Eq. (65) of Ref. \cite{covinst}).  This property implies that the output of the isometric channel  $\map V (\rho) = V \rho V^\dag$ has support in the subspace  $\left(\spc K^{\otimes M}\right)_+ \otimes \spc H_{in}$, where $\spc K = \spc H^{\otimes 2}$.  Now, consider the channel $\map F = \Tr_{in} \circ \map V:  {\sf Lin} (\spc H_{in}) \to {\sf Lin} \left( (\spc K^{\otimes M}  )_+\right)$.    By theorem \ref{theo:ssymbroad}, there exists a measure-and-prepare channel  $\widetilde {\map F}$ of the form $\widetilde{\map F} (\rho) = \sum_{i}  \Tr[P_i \rho] ~ |\Psi_i\>\<\Psi_i|^{\otimes M}$, with $|\Psi_i\> \in \spc H^{\otimes 2}$,  such that  the restrictions $\map F^{(k)}$ and $\widetilde{\map F}^{(k)}$ satisfy the bound of Eq. (\ref{bound3}) with the substitution $d \to d^2$.  To obtain the desired result it is sufficient to define the channel $\widetilde {\map E}$ as  $ \widetilde{\map E} (\rho) =    \Tr_{env} [\widetilde{\map V} (\rho) ]= \sum_i \Tr[P_i \rho]  ~\rho_i^{\otimes M}$, where $\rho_i$ is the reduced density matrix of $|\Psi_i\>\<\Psi_i|$, and  to use the  relation 
\begin{equation*}
\begin{split}
\|  \widetilde {\map E}^{(k)} - \map E^{(k)} \|_\diamond &= \|  \Tr_{env,k} \circ (\widetilde{\map F}^{(k)} - \map F^{(k)})\|_\diamond\\
& \le  \|  \widetilde{\map F}^{(k)} - \map F^{(k)}\|_\diamond,
\end{split}
\end{equation*}
 where $\Tr_{env,k}$ denotes the partial trace over the $k$ systems in the environment. \qed 

\emph{Observation.} The usual de Finetti theorems for quantum states \cite{oneandhalf,renner,facile} can be retrieved from theorems \ref{theo:ssymbroad}  and \ref{theo:symbroad} in the special case of symmetric broadcasting channels with trivial input space $\spc H_{in} \simeq \Cmplx$. In this case the POVM $\{P_i\}$ becomes just a collection of probabilities $\{p_i\}$.  

\medskip Theorems \ref{theo:ssymbroad} and \ref{theo:symbroad} have many interesting consequences: first of all they imply that the output state of $k$ receivers contains a vanishing amount of entanglement in the limit of vanishing $k/M$. Moreover, they imply that the information transmitted to a small number of receivers can only be classical, while the amount of quantum information 
is vanishing.  This observation will be made quantitatively precise in subsection \ref{sec:capacitybounds}. Another consequence is a strong form of the  equivalence between asymptotic cloning states estimation, briefly discussed in the next subsection.

\subsection{Strong equivalence between  asymptotic pure state cloning and state estimation}\label{subsec:strongeq}
Let $\{|\psi_x\>\}_{x\in X} \subset \spc H$ be a set  of pure states and $\{p_x\}$ a corresponding set of prior probabilities.  An $N$-to-$M$ cloning channel transforms $N$ copies of a state $|\psi_x\>$  into $M$ approximate copies, the joint state of the copies being a state on $\spc H^{\otimes M}$.  The requirement that each single copy  have the same fidelity with the state $|\psi_x \>$ is implemented without loss of generality by taking cloning channels with permutationally invariant output: clearly, such cloning channels are an example of symmetric broadcast channels.  Let us call $\map {Clon}_{N,M}$ the $N$-to-$M$ cloning channel under consideration and let $\widetilde {\map {Clon}}_{N,M}$ be the measure-and-prepare channel defined in Theorem \ref{theo:symbroad}.  Theorem \ref{theo:symbroad} then implies the bound
\begin{equation}\label{convergencetoest}
\left| \! \left|    \map{Clon}_{N,M}^{(k)}  -\widetilde {\map Clon}_{N,M}^{(k)}  \right| \! \right|_\diamond \le \frac{2 d^2 k}M,
\end{equation}  
that is, for fixed $k$ and $d$ the cloning channel becomes more and more indistinguishable from a measure-and-prepare channel as $M$ increases.  In particular, if $\map {Clon}_{N,M}$ is the optimal cloning channel according to some figure of merit, Eq. (\ref{convergencetoest}) entails the convergence of optimal cloning to estimation.  Note that the convergence in diamond norm represents an improvement over the trace-norm convergence of Ref. \cite{me}, as it states that cloning is indistinguishable from estimation even with the aid of entanglement with a reference system. 
The convergence of the fidelities is then a simple corollary: For every state $\psi_x$, the single-copy fidelity is  given by 
\begin{equation*}
F_{clon} [N,M,x] =  \<\psi_x|    \map {Clon}^{(1)}_{N,M}  (\|\psi_x\>\<\psi_x|^{\otimes M}) |\psi_x\>.
\end{equation*}
Denoting by $F_{\widetilde {clon}} [N,x]$ the single-copy fidelity for the measure-and-prepare channel $\widetilde{\map{Clon}}_{N,M}$ (note that in this case the fidelity is independent of $M$), we have 
\begin{equation*}
\begin{split}
|F_{clon}[N,M,x] - F_{\widetilde{clon}}[N,x] &|\le \left|\!\left|  (\map{Clon}^{(1)}_{N,M}  -\widetilde{\map {Clon}}^{(1)}_{N,M}) (|\psi_x\>\<\psi_x|^{\otimes N})  \right|\!\right|_1\\
&   \le \left|\!\left|  \map{Clon}^{(1)}_{N,M}  -\widetilde{\map {Clon}}^{(1)}_{N,M}  \right|\!\right|_\diamond \le \frac{2d^2 k} M.
\end{split}
\end{equation*}
Denoting by $F_{est} [N]$ the maximum average fidelity achievable by a measure-and-prepare channel and using the fact that $F_{est}[N] \le  F_{clon}[N,M], \forall M$ we then have the bound
\begin{equation*}
\begin{split}
0 \le F_{clon}[N,M] - F_{est}[N]  &\le  \left| \sum_{x}  p_x  (F_{clon}[N,M,x] - F_{\widetilde {clon}}[N,x])  \right| \\
&\le  \sum_{x}  p_x  \left| F_{clon}[N,M,x] - F_{\widetilde {clon}}[N,x] \right|  \le  \frac{2d^2 k}M,
\end{split}
\end{equation*}
which implies the limit $\lim_{M\to \infty} F_{clon} [N,M] = F_{est}[N]$.    

\subsection{Bounds on the quantum capacities of the $k$-receivers restriction of a symmetric broadcast channel}\label{sec:capacitybounds}

Theorems \ref{theo:ssymbroad}  and \ref{theo:symbroad}  also imply a set of bounds on the amount of quantum information that $k$ receivers can jointly decode  at the output of a symmetric broadcast channel $\map E$.   For definiteness, let us consider the case of a channel $\map E$ with output in the symmetric subspace $\left( \spc H^{\otimes M} \right)_+$: this is the case, e.g. of all known examples of optimal pure state cloning \cite{clonrev}.  A first bound on the quantum capacity comes from the continuity result of Ref.\cite{debbiesmith}, that, along with the fact that measure-and-prepare channels have zero quantum capacity, yields the following estimate
\begin{equation}\label{qbound1}
Q (\map E^{ (k) }) = |Q( \map E^{(k)}) - Q( \widetilde{\map E}^{ (k) }) | \le \frac{16 kd} M \log d_+^{(k)} +4H\left(\frac{2kd} M\right).
\end{equation}
where  $H$ is the binary entropy $H(x)= -x\log x - (1-x) \log(1-x)$, and $\log$ denotes the logarithm in base 2.  

Two other estimates are given  in the following
\begin{corollary}
The quantum capacity of the $k$-receivers restriction of a symmetric broadcast channel $\map E: {\sf Lin} (\spc H_{in}) \to {\sf Lin} \left(  ( \spc H^{\otimes M})_+ \right)$ satisfies the bound
\begin{eqnarray}\label{qbound2}
Q(\map E^{(k)}) &\le&  \min\left\{  \log\left( 1 +\frac{2kd d_+^{(k)}  } M\right),\log\left( 1 +\frac{2kd d_{in} } M\right) \right\}\\
&\le & \min \left\{  \frac{2kd d_+^{(k)}  } M , \frac{2kd d_{in} } M \right\}\label{linearized}
\end{eqnarray}
\end{corollary}
{\bf Proof}   Holevo and Werner  proved that the quantum
capacity of a channel $\map C$ is upper bounded by the $\varepsilon$-quantum
capacity $Q_\varepsilon (\map C)$ \cite{holevowerner} (i.e. the supremum of the rates that are asymptotically achievable with error bounded by $\varepsilon$), and that $Q_\varepsilon (\map C)$ is upper bounded by $\log \| \map C {\rm\Theta}_{in}\|_\diamond$, where
${\rm \Theta}_{in}$ is the transposition map on the input space $\spc
H_{in}$.  We then obtain
\begin{equation*}
\begin{split}
  Q (\map E^{(k)}) \le Q_\epsilon (\map E^{(k)}) &\le \log  \| \widetilde{\map E}^{(k)} {\rm \Theta_{in}}+ (\map E^{(k)}-\widetilde {\map E}^{(k)}){\rm \Theta}_{in} \|_\diamond \\
  & \le  \log \left(      \| \widetilde{\map E}^{(k)} {\rm \Theta_{in}}\|_\diamond+ \| \map E^{(k)}-\widetilde {\map E}^{(k)} \|_\diamond \| {\rm \Theta}_{in} \|_\diamond \right)\\
  &\le \log\left( 1 +\frac{2kd d_{in} } M\right),
\end{split}
\end{equation*}
having used  the  triangle inequality, the submultiplicativity $|\!|  \map A \map B |\!|_\diamond \le |\!|  \map A |\!|_\diamond  |\!| \map B |\!|_{\diamond}$ the fact that $\| \widetilde{\map E}^{(k)} {\rm \Theta}_{in}\|_\diamond = 1$ since  $\widetilde{\map E}^{(k)} {\rm \Theta}_{in}
(\rho)=\int \d \varphi \Tr[ Q^T_\varphi \rho]
|\varphi\>\<\varphi|^{\otimes k}$ is still a quantum channel,  the equality $\| {\rm \Theta}_{in} \|_\diamond = d_{in}$, and the bound of Eq. \eqref{bound3}.  Similarly, denoting by  ${\rm \Theta}_+^{(M)}$ and ${\rm \Theta}_+^{(k)}$ the transposition maps on $\left(\spc H^{\otimes M}\right)_+$ and $\left(\spc H^{\otimes k}\right)_+$, respectively, we obtain  
\begin{equation*}
\begin{split}
  Q (\map E^{(k)}) \le Q_\epsilon (\map E^{(k)}) &\le \log  \| \widetilde{\map E}^{(k)} {\rm \Theta}+ (\map E^{(k)}-\widetilde {\map E}^{(k)}){\rm \Theta}_{in} \|_\diamond\\
  &\le \log\left[ 1 + \| (\map{UMeasPrep}_{M,k} -\Tr_{M-k}) {\rm
      \Theta}^{(M)}_+  ({\rm \Theta}^{(M)}_+ \map E {\rm
      \Theta}_{in})\|_{\diamond} \right]\\
 &\le \log\left[ 1 + \|  (\map {UMeasPrep}_{M,k} - \Tr_{M-k}){\rm
      \Theta}^{(M)}_+\|_{\diamond}
  \right]\\\
   &\le \log\left[ 1 + \| {\rm \Theta}_+^{(k)} \|_\diamond \| {\rm \Theta}_+^{(k)}  (\map {UMeasPrep}_{M,k} - \Tr_{M-k}){\rm
      \Theta}^{(M)}_+\|_{\diamond}
  \right]\\\
&= \log\left[ 1 + \| {\rm \Theta}_+^{(k)} \|_\diamond \| \map {UMeasPrep}_{M,k} - \Tr_{M-k} \|_{\diamond}
  \right]\\\
  & \le   \log\left( 1 +\frac{2kd d_+^{(k)}  } M\right).
\end{split}
\end{equation*}
having used the triangle inequality,  the  submultiplicativity $|\!|  \map A \map B |\!|_\diamond \le |\!|  \map A |\!|_\diamond  |\!| \map B |\!|_{\diamond}$, the fact that ${\rm \Theta}^{(M)}_+ \map E {\rm \Theta}_{in}$ is a channel and that ${\rm  \Theta}^{(k)}_+ \map{UMeasPrep}_{M,k}  {\rm \Theta}^{(M)}_+ = \map {UMeasPrep}_{M,k}$ and  ${\rm \Theta}_+^{(k)}  \Tr_{M-k} {\rm  \Theta}_+^{(M)} = \Tr_{M-k}$. The two bounds above prove Eq. (\ref{qbound2}). Eq. (\ref{linearized}) then follows immediately from the relation $\log (1+x) \le x$. \qed
Since the input quantum information has to be spread uniformly over a large number of receivers, a finite group of $k<<M$ receivers can only access a vanishing amount of information.  This fact holds even if the overall channel $\map E$ is unitary (for example, if $\map E$ is the identity channel from a super-user holding all input systems to $M$ users, each of them receiving one output system).  
 
\section{Conclusions}
In this paper we have seen that the standard finite quantum de Finetti theorems can be naturally rephrased as theorems about the diamond-norm distance between the optimal measure-and-prepare channel from $M$ to $k$ copies and the trace channel $\Tr_{M-k}$.  The working principle of the theorems appears to be the simple fact that estimation and re-preparation from $M$ to $k$ copies becomes almost ideal whenever $M$ is large with respect to $k$.  This idea suggests that similar approximation theorems could be obtained from other measure-and-prepare protocols based on estimation, where the input is given by $M$ copies of some state $|\psi_x\>, x \in X$ and the goal is to produce $k$ approximate copies. In this case, one can expect to obtain approximation theorems for multipartite quantum states in the linear span of the projectors $|\psi_x\>\<\psi_x|^{\otimes M}$.   The exploration of such generalizations is an interesting direction of future research.


\bigskip 
{\bf Acknowledgements.} 
I would like to thank D. Gottesman, R. Spekkens, I. Marvian, and A. Harrow for stimulating questions that helped me to improve the presentation.   
 Research at Perimeter Institute is supported by the Government of Canada through Industry Canada and by the Province of Ontario through the Ministry of Research and Innovation.

\section*{Appendix}

The Appendix is devoted to the derivation of Eq. (\ref{main}).   To this purpose we will use the
fact that every operator $\rho\in {\sf Lin} \left( (\spc H^{\otimes M})_+ \right)$  can be written as a
linear combination of the rank-one projectors $|\psi\>\<\psi|^{\otimes
  M}$.   An easy proof of this fact is given as follows: Let us write $|\psi\> = \sum_{k=1}^d  \psi_k |k\>$. Then, we have (cf. Eq. 2 of Ref. \cite{wernerclon})
  \begin{equation*}
  |\psi\>^{\otimes M}  = \sum_{\vec n \in\mathcal P_{M,d}}    \psi_1^{n_1}  \dots \psi_d^{n_d}     \sqrt{\frac{  M!} {n_1! \dots n_d!} }  |\vec n\>,  \end{equation*} 
  and also 
  \begin{equation*}
  \frac 1 {M! } \left.   \left( \prod_{k=1}^d  \frac{1}{\sqrt{m_k!}}  \frac{    \partial^{m_k}}{ \partial \psi_k^{m_k}}  \right) \left( \prod_{l=1}^d \frac 1 {\sqrt{n_l !}} \frac{  \partial^{n_l}}{ \partial {\psi}_k^{*n_l}} \right)  |\psi \>\<\psi|^{\otimes M}  \right |_{\psi = 0} =|\vec m \>\<\vec n| ,  
  \end{equation*}  
  where the coefficients $\{\psi_k\}_{k=1}^d$ and their complex conjugates $\{\psi_l^*\}_{l=1}^d$ are treated as independent variables. 
  This means that the operators $|\vec m\>\<\vec n|$ are in the linear span of the projectors $|\psi\>\<\psi|^{\otimes M}$ (indeed, the derivatives are limits of linear combinations, and, since we are in finite dimensions, any linear span is a closed set, containing all its limit points).  Since the operators $\{|\vec m\>\<\vec n|\}_{\vec m, \vec n \in \mathcal P_{M,d}}$  span $\mathsf{Lin} \left(\left( \spc H^{\otimes M} \right)_+\right)$, the projectors $|\psi\>\<\psi|^{\otimes M}$ also do.  Note that the same conclusion would be obtained, through a lengthier calculation,  by taking all possible derivatives with respect to the real parts $\{  \mathsf{Re} (\psi_k)\}_{k=1}^d$ and the imaginary parts $\{\mathsf{Im} (\psi_k)\}_{k=1}^d$, instead of the derivatives with respect to the coefficients $\{\psi_k\}_{k=1}^d$ and their complex conjugates $\{\psi_k^*\}_{k=1}^d$.    
  
  Due to the above discussion, to prove Eq. (\ref{main}) it is enough to characterize the action of $\map
{UMeasPrep}_{M,k}$ on a generic projector $|\psi\>\<\psi|^{\otimes M}$.
Moreover, since the choice of the basis $\{|1\>,|2\>, \dots, |d\>\}$
is arbitrary, for given $|\psi\>$ we can choose $|1\> =
|\psi\>$.  Then, Eq. (\ref{canaleproiettore}) gives  
\begin{equation*}
\map{UMeasPrep}_{M,k} (|1\>\<1|^{\otimes M}) = \frac{d^{(M)}_+ }{d^{(M+k)}_+}
\sum_{ \vec m, \vec n\in \spc P_{k,d}} \alpha_{\vec m,  \vec n} |\vec
m\>\< \vec n| 
\end{equation*}  
with $\alpha_{\vec m, \vec n} = \< 1|^{\otimes M}
\<\vec m | P^{(M+k)}_+ |1\>^{\otimes M} |\vec n\>$.
Using the relation 
\begin{equation*}
P^{(M+k)}_+ = \frac 1 {(M+k)!}  \sum_{\pi \in S_{M+k}}
U^{(M+k)}_\pi
\end{equation*}
and Eq. (\ref{symbasis}) with the substitution $M\to k$, we  obtain  $ \alpha_{ \vec m, \vec n} =\frac{  k!(M+ n_1 )!} {(M+k)! n_1!} \delta_{\vec m, \vec n} $, and, therefore,
\begin{equation}\label{first}
 \map {UMeasPrep}_{M,k} (|1\>\<1|^{\otimes M}) = \frac{d_+^{(M)}}{d_+^{(M+k)}} \begin{pmatrix} M+k \\ k \end{pmatrix}^{-1}  \sum_{\vec n \in \spc P_{k,d}}  \begin{pmatrix} M + n_1  \\ M \end{pmatrix} |\vec n\>\<\vec n|.
\end{equation}

Using again  Eq. (\ref{symbasis}) with the substitution $M \to k$ we get the chain of equalities
\begin{equation*}\label{catena}
\begin{split}
 & \sum_{\vec n \in \spc P_{k,d}} \begin{pmatrix} M+ n_1  \\ M
  \end{pmatrix} |\vec n\>\<\vec n| =\\
&=\sum_{\vec n \in \spc P_{k,d}}
\left(  \frac{
\begin{pmatrix} M + n_1\\ M
\end{pmatrix}
} {k! n_1! \dots n_d!}  \sum_{\pi,\sigma \in S_k} U^{(k)}_\pi
\left(|1\>\<1|^{\otimes n_1} \otimes \dots \otimes |d\>\<d|^{\otimes n_d} \right)
U^{(k)}_\sigma\right) \\
&= \sum_{n_1 =0}^k 
\frac{
  \begin{pmatrix} M+ n_1  \\ M
  \end{pmatrix}
} {k! n_1! (k-n_1)!}  \sum_{\pi,\sigma \in S_k} U^{(k)}_\pi
\left(|1\>\<1|^{\otimes n_1} \otimes (I-|1\>\<1|)^{\otimes (k-n_1)} \right)
U^{(k)}_\sigma\\
&= \sum_{n_1=0}^{k} \sum_{j=0}^{k-n_1} 
\frac{ (-1)^{j}
  \begin{pmatrix} M+ n_1  \\ M
  \end{pmatrix}
\begin{pmatrix} k-n_1\\ j
  \end{pmatrix}
} {k! n_1! (k-n_1)!}  \sum_{\pi,\sigma \in S_k} U^{(k)}_\pi
\left(|1\>\<1|^{\otimes (n_1 + j)} \otimes I^{\otimes (k-n_1-j)}
\right) U^{(k)}_\sigma.
\end{split}
\end{equation*}
Defining $s= n_1 + j$, the chain can be continued as
\begin{equation*}
\begin{split}
 & \sum_{\vec n \in \spc P_{k,d}} \begin{pmatrix} M+n_1  \\ M
  \end{pmatrix} |\vec n\>\<\vec n| =\\
&= \sum_{n_1 =0}^k\sum_{s=n_1}^{k} 
\frac{ (-1)^{s-n_1}
  \begin{pmatrix} M + n_1  \\ M
  \end{pmatrix}
\begin{pmatrix} k-n_1\\ s-n_1
  \end{pmatrix}
} {k! n_1! (k-n_1)!}  \sum_{\pi,\sigma \in S_k} U^{(k)}_\pi
\left(|1\>\<1|^{\otimes s} \otimes I^{\otimes (k-s)}
\right) U^{(k)}_\sigma\\
&= \sum_{s =0}^k\sum_{n_1=0}^{s} 
(-1)^{s-n_1}
  \begin{pmatrix} M + n_1  \\ M
  \end{pmatrix}
\begin{pmatrix} k\\ s
  \end{pmatrix}
\begin{pmatrix}
s\\n_1
\end{pmatrix}  
 ~ P^{(k)}_+
\left(|1\>\<1|^{\otimes s} \otimes I^{\otimes (k-s)}
\right) P^{(k)}_+\\
\end{split}
\end{equation*}
Finally, we can use the combinatorial identity (see  proof below)
\begin{equation}\label{betas}
\beta_s := \sum_{n =0}^s  (-1)^{s-n}  \begin{pmatrix}    s \\ n\end{pmatrix} \begin{pmatrix}    M+n \\ M\end{pmatrix}  =  \begin{pmatrix}    M\\ s\end{pmatrix} 
\end{equation} 
 to obtain 
\begin{equation}\label{last}
  \sum_{\vec n \in \spc P_{k,d}} \begin{pmatrix} M + n_1  \\ M
  \end{pmatrix} |\vec n\>\<\vec n|   = \sum_{s=0}^k \begin{pmatrix} k \\s \end{pmatrix}\begin{pmatrix} M
  \\ s \end{pmatrix} 
  ~ P^{(k)}_+ (|1\>\<1|^{\otimes s} \otimes I^{\otimes k-s}) P^{(k)}_+.
\end{equation} 
Since $\begin{pmatrix}   M \\s\end{pmatrix} = 0$ whenever $s >M$, the sum is in fact a sum from $0$ to $\min\{M,k\}$.
Combining Eqs. (\ref{first}), (\ref{last}), and (\ref{clon})  we obtain the expression
\begin{equation*}\label{finalissima}
\begin{split}
\map{UMeasPrep}_{M,k} (|1\>\<1|^{\otimes M}) &=    
\sum_{s=0}^{\min \{k.M\} } \frac{d^{(M)}_+ \begin{pmatrix} k \\s \end{pmatrix}\begin{pmatrix} M
  \\ s \end{pmatrix}}{d^{(M+k)}_+ \begin{pmatrix}  M+k \\ k \end{pmatrix}} ~P^{(k)}_+ (|1\>\<1|^{\otimes s} \otimes
I^{\otimes k-s}) P^{(k)}_+\\ 
&   =  \sum_{s=0}^{\min\{k,M\}}  \frac{\begin{pmatrix}  M \\ s \end{pmatrix} \begin{pmatrix}  d+k-1 \\ k-s \end{pmatrix}}{\begin{pmatrix}  d+M +k-1 \\ k \end{pmatrix}} ~ \map {UClon}_{s,k} \left( |1\>\<1|^{\otimes s}\right), 
\end{split}
\end{equation*}
which holds for arbitrary $M$ and $k$, and for an arbitrary vector
$|1\> = |\psi\>$.  Hence, we  have obtained Eq.
(\ref{main}).  

  
Regarding the combinatorial identity of Eq. (\ref{betas}), it can be proved as follows: First, using Chu-Vandermonde formula (Eq. (\ref{chuvander})) one obtains 
$\beta_s =  \sum_{n=0}^s  \sum_{l=0}^M  (-1)^{s-n}  \begin{pmatrix}    s \\ n\end{pmatrix}  \begin{pmatrix}    s+n \\ l\end{pmatrix} \begin{pmatrix}    M-s \\ M-l\end{pmatrix}  $    
Then, Klee's identity (Proposition 1.1 of Ref. \cite{Klee}) yields $\beta_s =\sum_{l=0}^{M}  \begin{pmatrix}    s \\ l-s  \end{pmatrix} \begin{pmatrix}    M-s \\ M-l\end{pmatrix} = \sum_{l' = 0}^{M-s}   \begin{pmatrix}    s \\ l'\end{pmatrix}  \begin{pmatrix}   M-s \\ M-s-l' \end{pmatrix} $. 
Finally, the expression $\beta_s  =      \begin{pmatrix}    M \\ s\end{pmatrix} $ follows by applying Chu-Vandermonde formula again.

\end{document}